\documentclass{iau} 
\usepackage{graphicx}

\title[Dust evolution: going beyond the empirical] 
{Dust evolution: going beyond the empirical}

\author[Nathalie Ysard]   
{Nathalie Ysard$^1$}

\affiliation{$^1$Institut d'Astrophysique Spatiale, CNRS, Univ. Paris-Sud, \\
Universit\'e Paris Saclay, B\^atiment 121, 91405 Orsay Cedex, France \\ email: {\tt nathalie.ysard@ias.u-psud.fr} \\[\affilskip]}

\pubyear{2019}
\volume{350}  
\jname{Laboratory astrophysics: from observations to interpretation}
\editors{Farid Salama, Helen Jane Fraser \& Harold Linnartz, eds.}
\begin{document}

\maketitle

\begin{abstract}
A key element when modeling dust in any astrophysical environment is a self-consistent treatment of the evolution of the dust material properties (size distribution, chemical composition and structure) as they react to and adjust to the local radiation field intensity and hardness and to the gas density and dynamics. The best way to achieve this goal is to anchore as many model parameters as possible to laboratory data. In this paper, I present two examples to illustrate how outstanding questions in dust modeling have been/are being moved forward by recent advances in laboratory astrophysics and what laboratory data are still needed to further advance dust evolution models.
\end{abstract}

\firstsection

\section{Introduction}

Interstellar grains are a major actor in the chemical, physical, and dynamical evolution of matter in the cosmos. Their emission dominates the observations of galaxies over the mid-IR to submillimetre electromagnetic spectrum (\cite{Galliano2018}) and should therefore be one of the most powerful diagnostic tools to trace processes at the heart of issues central to contemporary astrophysics and in particular those leading to the formation of stars and planets. However, this power is compromised by gaps in our knowledge of the exact properties of dust. Above all, it is now well established that dust is not the same everywhere (e.g. \cite{PCXI, Remy2018}): it is the evolution of grains in all its complexity that must be understood and modelled to allow an optimal use of the astronomical data. It is only very recently that interest in such a refined study of grains has gone beyond the theoretical stage. Indeed, with the advent of telescopes such as ALMA, VLTI-MATISSE, JWST, and SKA, we now (will) have access to data with unprecedented spatial and spectral resolutions. With these telescopes, the small transition zones between the phases of the interstellar medium (ISM), where most of the grain evolution takes place, are (will) be spatially resolved. In addition, technological progress now makes it possible to calculate the optical properties of very complex grains in a reasonable time. However, such a degree of refinement in dust models can only be achieved if as many parameters as possible are anchored to the data obtained by laboratory astrophysicists: pure and mixed material optical properties; impact of UV photons, cosmic rays,  and grain-grain collisions on dust composition, size and structure again in relation with their optical properties; etc. If not, the number of degrees of freedom of these models will always be too large for the interpretation of astronomical observations to lead to definitive conclusions on grain properties.

This paper is organized as follows. Section~\ref{observations} gives a short overview of the latest dust observational constraints in the Milky Way from diffuse to dense ISM and where they contradict with current dust model predictions. Section~\ref{laboratory} then presents two examples of recent laboratory experiments that are likely to resolve some of these contradictions in the near future. Finally, Sect.~\ref{conclusion} gives concluding remarks.

\section{Observational constraints vs. modeling}
\label{observations}

{\underline{\it Diffuse interstellar medium}}. Startling observations have recently sparked a renewed interest in the diffuse ISM from the community. This low brightness optically thin medium was thought to be rather homogeneous in terms of grain properties and more or less well explained by current dust models. That was before the advent of the Planck-HFI data, which shattered this image of a relatively simple medium. Confronting the Planck all-sky data with independent optical extinction estimates, the Planck Collaboration showed that the emission-to-extinction ratio predicted by standard grain models (\cite{Draine2007}, \cite{Compiegne2011}) is wrong by a factor up to 2 (\cite{Fanciullo2015}, \cite{PCXXIX}). This factor directly reflects upon dust mass estimates. Then, investigating diffuse lines-of-sight with $N_H < 3 \times 10^{20}$~H/cm$^2$, they further showed that the dust luminosity is independent of its temperature, whereas its opacity decreases with increasing temperature (\cite{PCXI}). This was totally counter-intuitive regarding what we thought we knew about diffuse ISM dust. Indeed, for a given set of dust properties, the luminosity should increase with temperature whereas the opacity should remain constant, this means that dust already evolves in the diffuse ISM (see also \cite{Siebenmorgen2018}). This evolution was confirmed by further studies, showing for instance an increase in the dust opacity by around 40\% when $N_H > 5 \times 10^{20}$~H/cm$^2$, in parallel with very little variations in $E(B-V) / N_H$ (\cite{NGuyen2018}). Variations in the submillimetre dust opacity along with variations in the gas-to-dust mass ratio were observed in individual diffuse regions and from cloud-to-cloud (\cite{Reach2017}), with one plausible explanation being that ``... grain properties may change within the clouds: they become more emissive when they are colder, while not utilizing heavy elements that already have their cosmic abundance fully locked into grains''. This means that the diffuse-ISM dust evolution would not be driven by silicate forming elements but rather by cycling of carbon in and out of dust. While significant changes in dust opacities were unexpected, significant variations in the gas-to-dust mass ratio were no less so. Until very recently, this ratio, referred to as the Bohlin's ratio (\cite{Bohlin1978}), was believed to be canonical. However, the latest observations showed that it is rather higher than previously thought by 20 to 60\% and subject to local variations (\cite{Liszt2014}, \cite{Lenz2017}, \cite{Murray2018}, \cite{NGuyen2018}). All these models vs. observations inconsistencies raise fundamental questions about the optical properties of grains that should be answered thanks to laboratory measurements.

In presence of a magnetic field, non-spherical grains tend to align with their main axis parallel to the field lines. Dichroic extinction of the radiation field by dust grains then results in polarization, parallel to the magnetic field in extinction and perpendicular to it in emission. The Planck telescope provided the first all-sky map of dust polarized emission in the submillimetre. Once again, the diffuse ISM observations do not match the predictions of the standard grain models (\cite{PCXXI}). Indeed the spectral shape of the polarization fraction in the submillimetre is flatter than expected and as for unpolarized data, the polarized extinction-to-emission ratio is higher by a factor of $\sim 2.5$ than model predictions (e.g. \cite{Draine2009}). This shows that existing dust models do not correctly describe either the dust optical properties or the mechanism of their alignment with the magnetic field.

Extinction observations have also recently brought their share of surprises. In particular, extinction in the mid-IR appeared to be flatter than predicted by current models (\cite{Wang2013}). This has been explained by invoking the presence of micron-sized graphitic grains (\cite{Wang2015}), far from the usual size distribution centered around 100~nm used to explain the dust thermal emission, polarized or not. However, such big grains would produce an additional thermal emission component conflicting with the Planck observations at $\sim 3$~mm. An alternative explanation is that the optical constants used to describe ISM grains need to be reassessed in the light of the newest laboratory measurements on analogues of ISM silicates (\cite{Demyk2017}, \cite{Mutschke2019}).

{\underline{\it Dense interstellar medium}}. Talking about ``the'' dense ISM is a bit tricky since this term covers a wide variety of regions with very different physical conditions. They do, however, have one factor in common: their grains are no longer isolated but have coagulated to form large aggregates. The growth of grains is most probably initiated by the accretion of gas phase elements and the coagulation of the grains with one another, in parallel or followed by ice mantle formation. In molecular clouds and denser cold cores, this growth was first evidenced by the decrease in the mid-IR to far-IR emission ratio (e.g. \cite{Stepnik2003}), showing the disappearance of the smallest grains. Growth is also evidenced by a decrease in the dust temperature from about 20~K in the diffuse ISM to less than 15~K in denser regions (e.g. \cite{Juvela2011}) along with an increase in the dust far-IR/submillimetre opacity and spectral index (e.g. \cite{Roy2003, Remy2018}). In the meantime, the total-to-selective extinction ratio $R_V$ significantly increases when $A_V > 3$ (e.g. \cite{Whittet2001, Campeggio2007}). Dust growth is also evidenced by an increased scattering efficiency from the near- to mid-IR when the local density increases (e.g. \cite{Foster2006, Pagani2010}). All these observations are in agreement with grain growth by coagulation to form aggregates. However, the models disagree as to the answer to the simplest a priori question: up to what size do the grains grow? Indeed, depending on how grains are modeled the answer can vary by up to one order of magnitude from $\sim 0.5$ to 5~$\mu$m (e.g. \cite{Steinacker2015, Lefevre2016, Ysard2016})
, which has a strong impact on the predictions of chemistry and dynamics simulations aimed at understanding the formation of prestellar cores from molecular clouds. The differences between the models come from the level of complexity integrated into the grain description in terms of its geometric structure and chemical composition.

Polarization observations, both in emission and extinction, also point towards variations in the grain properties when the ISM density increases (\cite{Fanciullo2017}). Moreover, these variations cannot just be changes in the grain size distribution but must encompass changes in grain shape, porosity, and chemical composition (\cite{Juvela2018}) in parallel with a loss in the efficiency of the grain alignement with the local magnetic field. Is this loss due to a decreased efficiency in the processes at the origin of grain alignement or to changes in the grain structure? Answering this question is of prime importance as it determines the level of aggregate coupling to the magnetic field as well as their surface-to-mass ratio, impacting thus both the gas dynamics and chemistry.

Beyond the significant increase in $R_V$ when $A_V > 3$ and variations in the mid-IR extinction curve, extinction data provide constraining information on the structure, composition and size of aggregates through the observation of ice and silicate bands (e.g. \cite{Dartois1998, Boogert2015}). Indeed, the exact band positions and shapes vary with the ice layer thickness, the grain size distribution and composition, and the aggregate structure. However, there are still many open questions. For instance, which happens first: grain-grain coagulation or ice layering? This determines what kind of materials are in contact in the aggregates and will most probably affect the band profiles. Similarly, we know that the shape of the individual grains composing the aggregates affects the silicate band profiles: is there an effect on the ice bands? The advent of the JWST should allow great advances in our understanding of ice properties at the scale of individual clouds and protoplanetary discs, however this will be possible only if prior laboratory experiments do give constraints on all these questions.

\section{Laboratory astrophysics inputs}
\label{laboratory}

Comparison of observational constraints with existing dust models clearly shows that much information from the laboratory is needed to advance our understanding of cosmic dust. Here, we concentrate on two examples that are probably soon going to change the way dust is modelled: (i) optical properties of amorphous silicates (e.g. \cite{Demyk2017}) ; (ii) dust size distribution (e.g. \cite{Lorek2018}).

{\underline{\it Optical properties: the example of amorphous silicates}}. Almost all dust models make use of silicate optical properties somehow derived from the semi-empirical astro-silicates defined by \cite{Draine1984} (see for instance \cite{Desert1990, Zubko2004, Compiegne2011, Jones2013, Siebenmorgen2014, Guillet2018}). Starting from olivine optical properties, the experimental dielectric function was modified to fit astronomical observations of the silicate vibration bands at $\sim 10$ and 18~$\mu$m from hot dust in the Trapezium region and from circumstellar dust. For wavelengths longer than 20~$\mu$m, the silicate emissivity was assumed to decrease in $\lambda^{-2}$. This semi-empirical model has been extremely useful as for years no laboratory data were available on a sufficiently large wavelength range. However, as described in the previous section, it was recently shown that, in the far-IR to submillimetre, the astro-silicates are not emissive enough and their spectral index is too steep, and that in the mid-IR their spectral features are too sharp. Recently, laboratory measurements on interstellar silicate analogues were performed from the mid-IR to the submillimetre by Demyk \etal\ (2017). The analogues were amorphous magnesium-rich glassy silicates with the stoichiometry of olivine and pyroxene (X35 and X50A/B samples, respectively, in Demyk \etal\ 2017). These measures of mass absorption coefficients (MACs) are of the utmost importance. Indeed, the measured bands at $\sim 10$ and 18~$\mu$m are much wider than those of astro-silicates (see Figs.~1 and 4 in Demyk \etal\ 2017), which is in line with recent astronomical observations (\cite{Wang2013}). Moreover, far from decreasing monotonically in $\lambda^{-2}$, the spectral index of the far-IR MACs varies as a function of wavelength, temperature, and analogue stoichiometry from $\sim \lambda^{-1}$ to $\lambda^{-3}$ (see Fig.~3 in Demyk \etal\ 2017).

\begin{figure}[!h]
\begin{center}
\begin{tabular}{cc}
\includegraphics[width=2.7in]{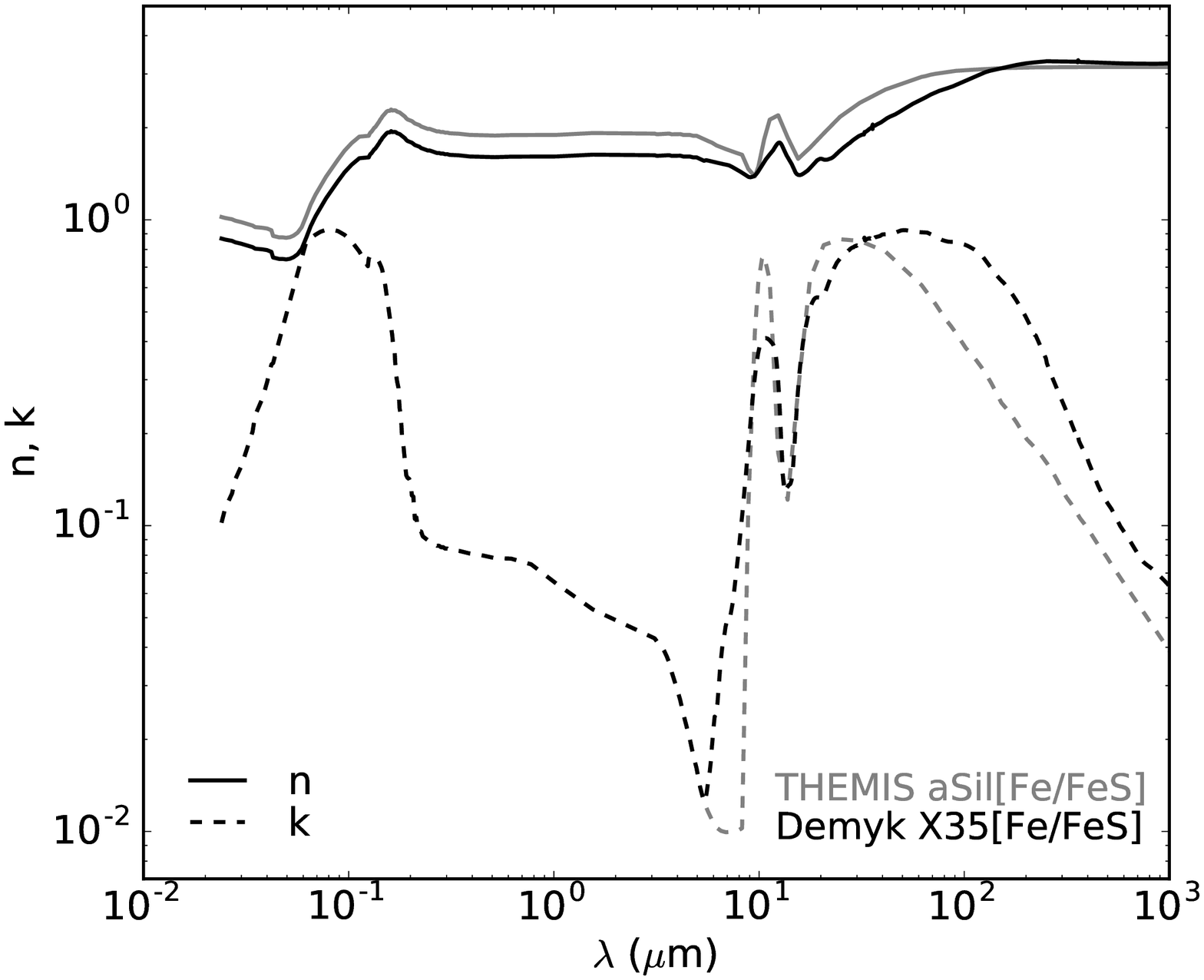} & \includegraphics[width=2.7in]{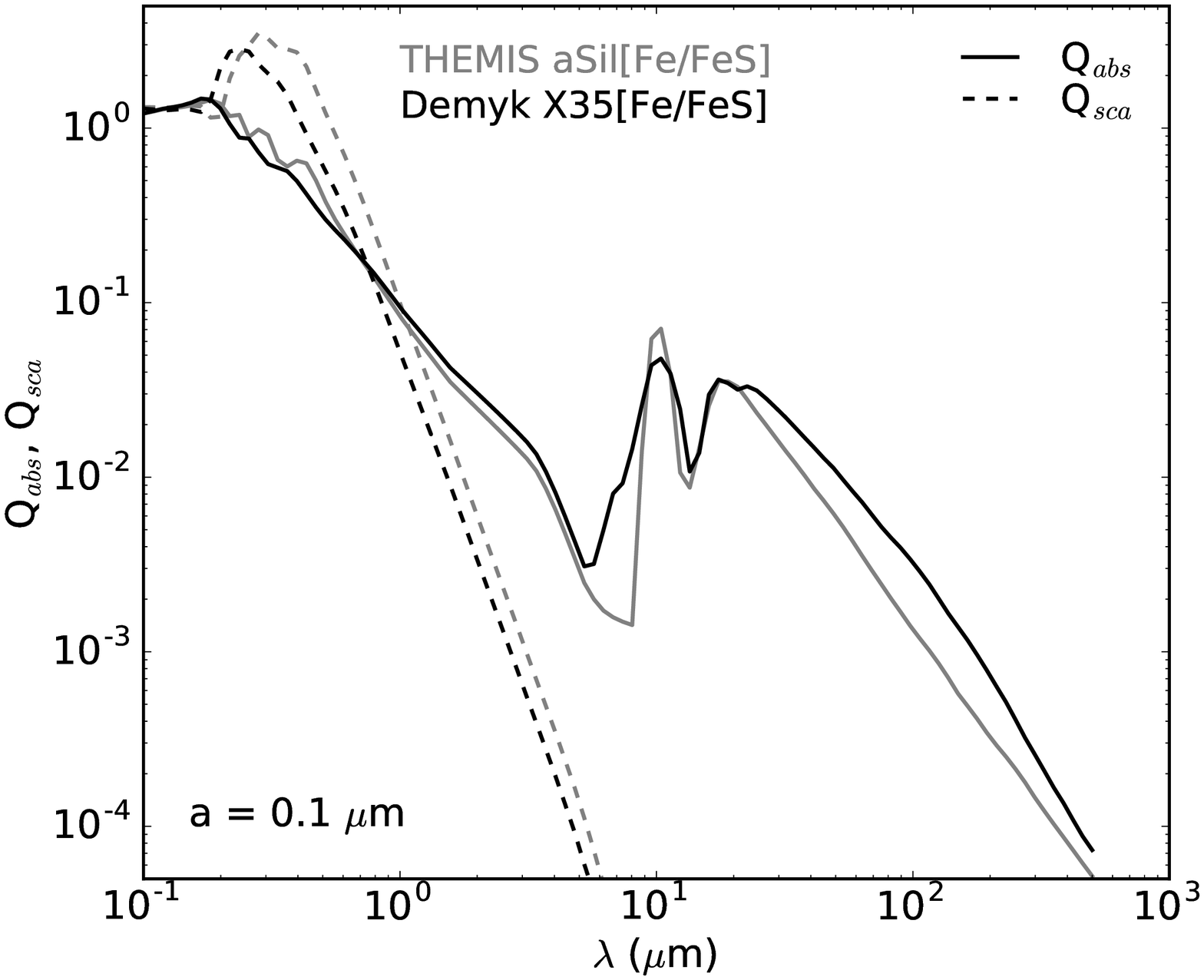}
\end{tabular}
\caption{{\it Left:} imaginary part $k$ (dashed lines) and real part $n$ (solid lines) of the complex refractive index. Gray lines show the THEMIS silicates with metallic nano-inclusions of Fe and FeS (see text for details and \cite{Koehler2014}). Black lines show the X35 sample considering the same metallic nano-inclusions as in THEMIS. {\it Right:} corresponding scattering efficiency $Q_{sca}$ (dashed lines) and absorption efficiency $Q_{abs}$ (solid lines) for a 0.1~$\mu$m sphere.}
\label{fig1}
\end{center}
\end{figure}

In order to compare these new laboratory measurements to the widely used astro-silicate model, a simplistic estimate of the complex refractive index $m = n + ik$ corresponding to the olivine X35 sample is presented in Fig.~\ref{fig1} (left figure). The X35 sample was chosen because it is a priori the least promising of the samples presented by Demyk \etal\ (2017) with a very steep MAC in the far-IR, a MAC which is also much lower than that of the other samples (see Fig.~4 in \cite{Demyk2017}). The calculations were performed assuming that the sample consists in an assembly of isolated 0.1~$\mu$m spheres and following the Kramers-Kronig relations (private communication from K. Demyk, A.P. Jones, V. Gromow, and C. M\'eny, paper currently in preparation). Starting from this new refractive index and using the Mie theory, the corresponding absorption and scattering efficiencies, $Q_{abs}$ and $Q_{sca}$ respectively, were estimated (right figure in Fig.~\ref{fig1}). To easily compare these results to the THEMIS model (The Heterogeneous dust Evolution Model for Interstellar Solids, Jones \etal\ 2013, Koehler \etal\ 2014), metallic nano-inclusions of Fe and FeS (10\% in volume) are included in the X35 grains. Three points are worth noting. First, as expected the X35 mid-IR spectral features are less sharp than those of the THEMIS version of astro-silicates. Second, the laboratory X35 silicate are twice as emissive for $80 \leqslant \lambda \leqslant 200~\mu$m and 25\% more emissive for $\lambda \geqslant 300~\mu$m. The emissivity should be even higher for the three other samples X40, X50A, and X50B presented in Demyk \etal\ (2017). Third, the X35 spectral index is flatter for $\lambda \leqslant 300~\mu$m. Looking at the MAC spectral shapes in Fig.~4 of Demyk \etal\ (2017), it is expected that this will be valid up to longer wavelengths for the other samples. These very preliminary results therefore show that switching from semi-empirical optical properties such as astro-silicates to optical properties anchored to laboratory measurements should a priori resolve the main contradictions between current dust models and astronomical observations of the diffuse ISM in extinction, total emission, and polarization.

\begin{figure}[!h]
\begin{center}
\begin{tabular}{c}
\includegraphics[width=5.5in]{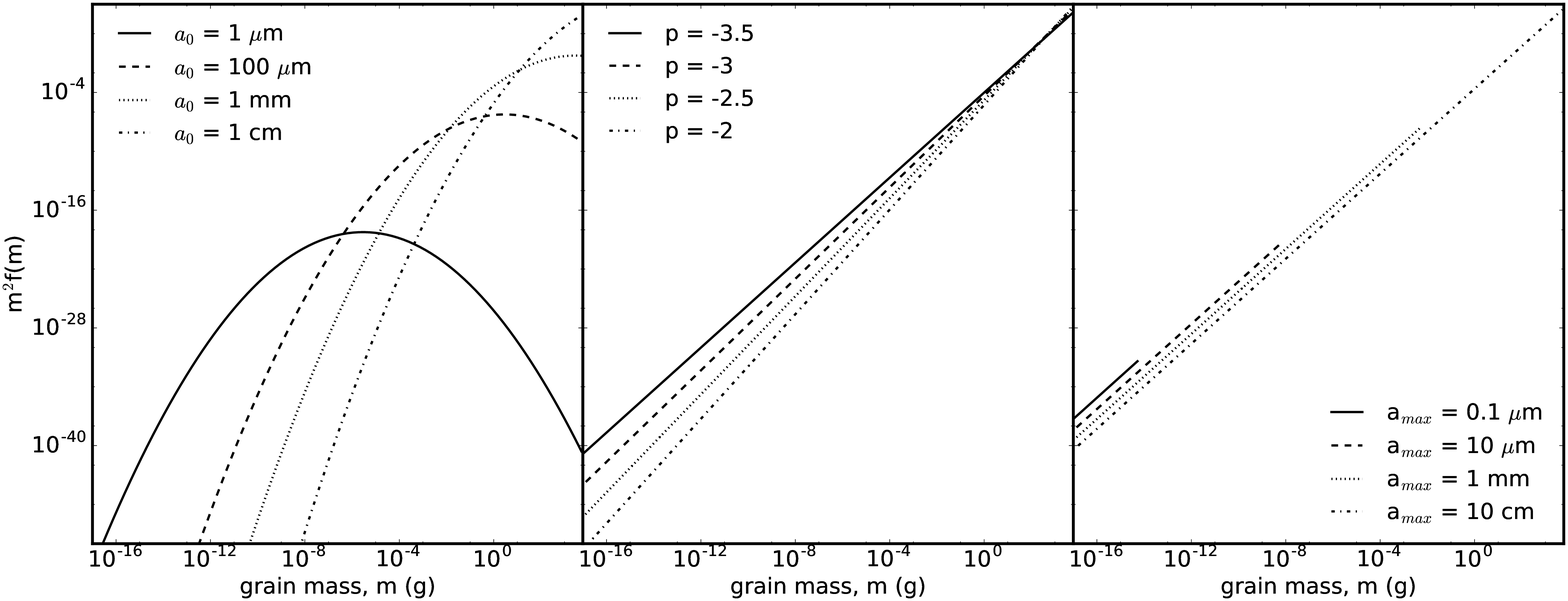} \\
\includegraphics[width=5.5in]{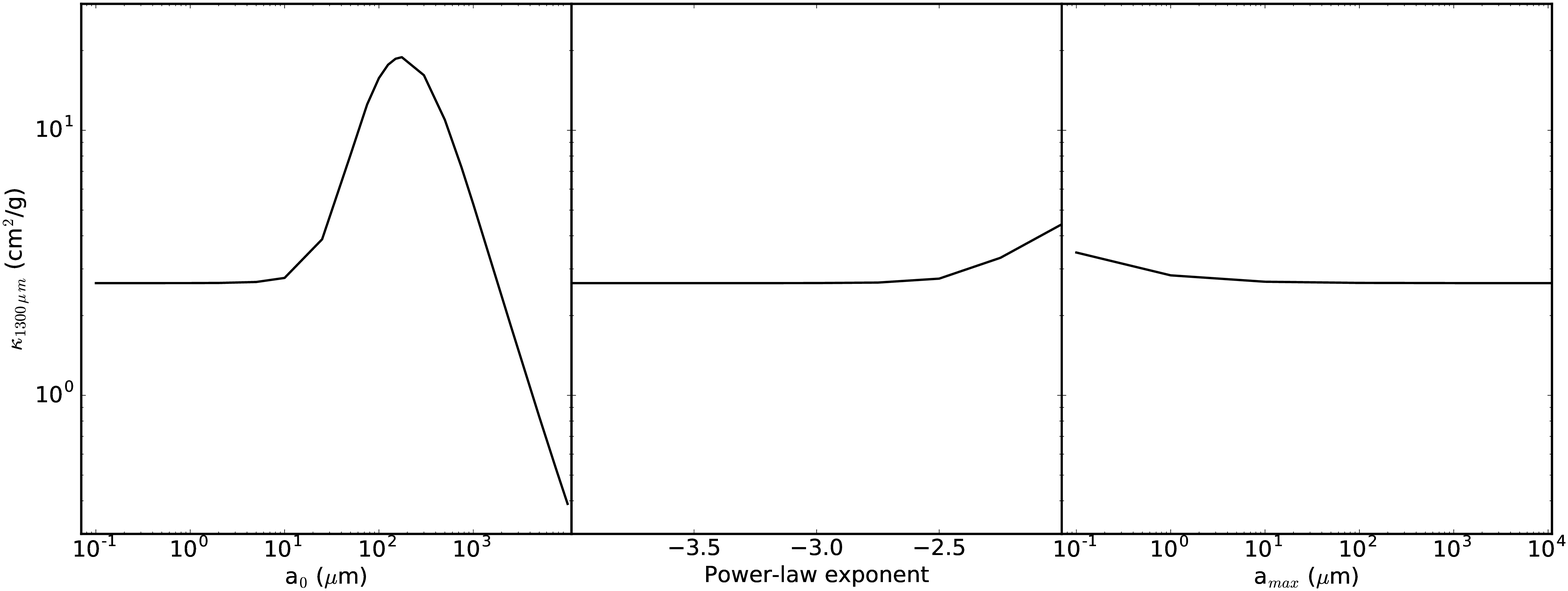} \\
\end{tabular}
\caption{{\it Top:} Mass distribution function $m^2f(m)$ per logarithmic mass bin, where $f(m)$ is the number of grains in the $[m, m+dm]$ mass interval, with a constant gas-to-dust mass ratio set at 100. The left figure shows log-normal size distributions for variable centroid sizes $a_0$ with $a_{min} = 0.01~\mu$m and $a_{max} = 10$~cm. The middle figure shows power-law size distributions for variable power-law exponents $p$ with $a_{min} = 0.01~\mu$m and $a_{max} = 10$~cm. The right figure also shows power-law size distributions but with a fix power-law exponent $p = -3.5$ and $a_{min} = 0.01~\mu$m, and variable $a_{max}$. {\it Bottom:} Corresponding mass absorption coefficients at 1.3~mm $\kappa_{{\rm 1.3~mm}}$ for log-normal size distributions as a function of $a_0$ (left), power-law size distributions as a function of $p$ (middle) and $a_{max}$ (right).}
\label{fig2}
\end{center}
\end{figure}

{\underline{\it Grain size distribution.}} An important point that comes out in many observational astrophysics studies is the question of the mass of the observed objects. A relatively common way to estimate it is to use grain continuum emission from the far-IR to the (sub)millimetre range and to assume that the dust mass absorption coefficient, $\kappa$, at a given wavelength is known: $\kappa = \int 3/4\rho \; dn/da \; Q_{abs}/a \; da$ for a given grain size distribution $dn/da$ and a grain volume density $\rho$. This was done, for example, for dense cloud cores by \cite{PCXXII} and for protostellar discs by \cite{Busquet2019}. However, many parameters are hidden behind the choice of an opacity as it depends on the grain size distribution, chemical composition, geometrical structure, and temperature distribution along the line-of-sight (itself dependent on the object geometry and illumination conditions). Let us focus on the influence of the size distribution on the opacity. The classical choice are power-law size distributions $dn/da \propto a^p$, where $p$ is most often set at -3.5 and sometimes varied, alone or in combination with the maximum grain size $a_{max}$ (see top middle and right figures in Fig.~\ref{fig2}), to account for grain growth in dense regions (e.g. \cite{Weidenschilling1977}, \cite{Natta2004}, \cite{Draine2006}). This type of size distribution is what is typically expected if grains are the result of collisional fragmentation cascades as in debris discs (e.g. \cite{Davis1990, Tanaka1996}). However, recent studies have shown that the grain size distribution during the collapse of dense prestellar cores and in protoplanetary discs can differ drastically from a power-law (e.g. \cite{Birnstiel2018}). For instance, Lorek \etal\ (2018) modelled grain growth in protoplanetary discs from a local perspective, focussing on grains ``... that would be concentrated in pebble-clouds by streaming instability and that would provide the building blocks of comets formed by the gravitational collapse of these clouds''. What makes this particular study interesting for us is that it is based on the latest laboratory results from \cite{Guttler2010} and \cite{Windmark2012} to predict the outcomes of grain-grain collisions (sticking, bouncing, fragmentation, erosion, mass transfer), from \cite{Guttler2010, Blum2008, Gundlach2011}, and \cite{Gundlach2015} for the sticking properties of water ice and silicate monomers, and from \cite{Guttler2009} and \cite{Weidling2009} to assume that not only compact grains but also aggregates can bounce. Considering various monomer sizes, radial positions in the disc, levels of turbulence, gas surface densities, etc., Lorek \etal\ (2018) always find size distributions differing quite notably from power-laws and that can be effectively modelled by log-normal type laws (see their Figs.~2 and 3 for instance and our top left figure in Fig.~\ref{fig2}): $dn/da \propto \exp \left[ -1/2 \left( \ln(a/a_0) / \sigma \right)^2 \right]$, where $a_0$ is the centroid of the size distribution and $\sigma$ its width.

Figure~\ref{fig2} presents the mass absorption coefficents $\kappa$ for power-law (varying $p$ or $a_{max}$) and log-normal size distributions (varying $a_0$) at 1.3~mm. They are calculated for porous spherical grains (porosity of 50\%) of which 2/3 of the volume is occupied by amorphous silicates containing metallic nano-inclusions of Fe and FeS and 1/3 by amorphous carbon. The optical properties are those of the THEMIS model. Huge differences between the two types of size distributions are found. The two power-law $\kappa$ show no variation when $-4 \leqslant p \leqslant 2.5$ and $1~\mu$m $\leqslant a_{max} \leqslant 10$~cm. Only when $p > -2.5$ or $a_{max} < 1~\mu$m increases of 1.7 and 1.3, respectively, are found. Variations with $a_0$ in the case of log-normal distributions are stronger. For $0.1 \leqslant a_0 \leqslant 10~\mu$m, $\kappa$ is almost constant, it then increases by a factor of $\sim 7$ when $a_0$ increases to $\sim 170~\mu$m, and decreases by a factor of $\sim 48$ when $a_0$ further increases to 1~cm (a factor of $\sim 7$ when compared to $a_0 = 1~\mu$m). This comes from the fact that, contrary to the power-law case, log-normal size distributions have a characteristic size $a_0$ which dominates the $\kappa$ estimate. According to the Mie theory, when $a \ll \lambda$, the increase in $Q_{abs}$ is proportional to $a$ leading to a constant $\kappa \propto Q_{abs} / a$. For larger sizes, but still smaller than the wavelength, the increase in $Q_{abs}$ is stronger, resulting in an increasing $\kappa$. For sizes comparable to or larger than the wavelength, $Q_{abs} \sim 1$ and is no longer dependent on the grain size, thus a decreasing $\kappa \propto 1/a$ with increasing $a_0$. The exact sizes at which the variations are observed as well as the level of increase or decrease in $\kappa$ depend on the chemical composition of the grains but the general trends with increasing $a_0$ remain the same (constant $\rightarrow$ increase $\rightarrow$ decrease). This illustrates how much the choice of a size distribution is crucial when modeling $\kappa$ to infer masses of ISM clouds or protoplanetary discs. Indeed, the variations in $\kappa$ presented here directly reflect on mass estimates. Laboratory experiments on grain growth are thus of the utmost importance if one wants to avoid the large uncertainties introduced by the choice of a size distribution in dust mass estimates.

\section{Conclusion}
\label{conclusion}

The quality of recent and future observations (e.g. Spitzer, Herschel, Planck, ALMA, SKA, JWST, Millimetron, SPICA...), in terms of spatial and spectral resolutions and sensitivity, means that grain models can no longer suffer from any approximation. Indeed, all existing dust models fail to explain all observations {\it simultaneously}. The best way to solve this problem is to anchor as many model parameters as possible to laboratory data even if this is far from being an easy task. The two examples presented earlier show how experiments to measure silicate optical properties and grain growth appear very promising to advance dust modelling. These are obviously not the only experiments the impact of which could be decisive for grain modelling. For instance, replicating the Demyk \etal\ (2017)'s experiments with amorphous carbon grains would probably be very instructive. In addition, laboratory experiments on the influence of grain shape on their phase function will be essential to understand self-scattering in protoplanetary discs. Experimental results on processing of grains by cosmic rays and UV photons should also be better incorporated into dust models (in terms of variations in size distribution and optical properties), for example to understand the future JWST observations of photo-dominated regions and protoplanetary discs. In this context, the latest laboratory measurements of PAHs and nano-carbon dust optical properties will be of prime importance. For larger grains, having the laboratory measured optical constants for various ice mantle compositions (e.g. amorphous water ice, mixtures of H$_2$0, CO and CO$_2$ ices) on a wide wavelength range would be a significant upgrading for dust models.


\end{document}